\documentclass[twocolumn]{aastex7}
\shorttitle{Bipolar ejecta flows in W\,49B}
\shortauthors{XRISM collaboration et al.}

\begin{document}

\title{Kinematic Evidence for Bipolar Ejecta Flows in the Galactic SNR W\,49B}

\correspondingauthor{Makoto Sawada}

\author[0000-0003-4721-034X]{Marc Audard}
\affiliation{Department of Astronomy, University of Geneva, Versoix CH-1290, Switzerland}
\email{Marc.Audard@unige.ch}
\author[0000-0001-7204-4350]{Hisamitsu Awaki}
\affiliation{Department of Physics, Ehime University, Ehime 790-8577, Japan}
\email{awaki@astro.phys.sci.ehime-u.ac.jp}
\author[0000-0002-1118-8470]{Ralf Ballhausen}
\affiliation{Department of Astronomy, University of Maryland, MD 20742, USA}
\affiliation{NASA/Goddard Space Flight Center, MD 20771, USA}
\affiliation{Center for Research and Exploration in Space Science and Technology, NASA/GSFC (CRESST II), MD 20771, USA}
\email{ballhaus@umd.edu}
\author[0000-0003-0890-4920]{Aya Bamba}
\affiliation{Department of Physics, University of Tokyo, Tokyo 113-0033, Japan}
\email{bamba@phys.s.u-tokyo.ac.jp}
\author[0000-0001-9735-4873]{Ehud Behar}
\affiliation{Department of Physics, Technion, Haifa 3200003, Israel}
\email{behar@physics.technion.ac.il}
\author[0000-0003-2704-599X]{Rozenn Boissay-Malaquin}
\affiliation{Center for Space Sciences and Technology, University of Maryland, Baltimore County (UMBC), MD 21250, USA}
\affiliation{NASA/Goddard Space Flight Center, MD 20771, USA}
\affiliation{Center for Research and Exploration in Space Science and Technology, NASA/GSFC (CRESST II), MD 20771, USA}
\email{rozennbm@umbc.edu}
\author[0000-0003-2663-1954]{Laura Brenneman}
\affiliation{Center for Astrophysics --- Harvard-Smithsonian, MA 02138, USA}
\email{lbrenneman@cfa.harvard.edu}
\author[0000-0001-6338-9445]{Gregory V. Brown}
\affiliation{Lawrence Livermore National Laboratory, CA 94550, USA}
\email{brown86@llnl.gov}
\author[0000-0002-5466-3817]{Lia Corrales}
\affiliation{Department of Astronomy, University of Michigan, MI 48109, USA}
\email{liac@umich.edu}
\author[0000-0001-8470-749X]{Elisa Costantini}
\affiliation{SRON Netherlands Institute for Space Research, 2333 CA Leiden, The Netherlands}
\email{e.costantini@sron.nl}
% ...

\collaboration{all}{XRISM Collaboration}

\begin{abstract}

W\,49B is a unique Galactic supernova remnant with centrally peaked, ``bar''-like ejecta distribution, which was once considered evidence for a hypernova origin that resulted in a bipolar ejection of the stellar core. However, chemical abundance measurements contradict this interpretation. Closely connected to the morphology of the ejecta is its velocity distribution, which provides critical details for understanding the explosion mechanism. We report the first-ever observational constraint on the kinematics of the ejecta in W\,49B using the Resolve microcalorimeter spectrometer on the X-ray Imaging and Spectroscopy Mission (XRISM). Using XRISM/Resolve, we measured the line-of-sight velocity traced by the Fe He$\alpha$ emission, which is the brightest feature in the Resolve spectrum, to vary by $\pm 300$~km~s$^{-1}$ with a smooth east-to-west gradient of a few tens of km~s$^{-1}$~pc$^{-1}$ along the major axis. Similar trends in the line-of-sight velocity structure were found for other Fe-group elements Cr and Mn, traced by the He$\alpha$ emission, and also for intermediate-mass elements Si, S, Ar, and Ca, traced by the Ly$\alpha$ emission. The discovery of the east-west gradient in the line-of-sight velocity, together with the absence of a twin-peaked line profile or enhanced broadening in the central region, clearly rejects the equatorially expanding disk model. In contrast, the observed velocity structure suggests bipolar flows reminiscent of a bipolar explosion scenario. An alternative scenario would be a collimation of the ejecta by an elongated cavity sculpted by bipolar stellar winds.

\end{abstract}

\section{Introduction} \label{sec:intro}

Spatial and velocity distributions of ejecta are key diagnostics used to probe the explosion mechanism and shock physics of supernova remnants (SNRs). Reverse-shocked layers of ejecta observed in young SNRs are believed to produce shell-like structure. There is also a growing number of so-called mixed-morphology SNRs \citep[MM SNRs:][]{Rho1998} in which centrally peaked X-rays fill radio-shell interiors. This peculiar structure triggered decades-long discussions, and attempts to theoretically model the morphology still continue today \citep{White1991, Shelton1999, Shimizu2012, Chiotellis2024}. The super-solar chemical abundances measured at the central region of some of MM SNRs indicate an ejecta origin for the interior of these remnants \citep[e.g.,][]{Bocchino2009}. Although the velocity structure would provide insight as to the origin of the morphology, there has been no robust imaging measurement of proper motion in MM SNRs. In addition, there has been no detection of significant Doppler shifts with X-ray CCD-based spectrometers. This may be because the ejecta speed is an order-of-magnitude slower, i.e., below the detection limit of the moderate energy resolution of CCDs, compared to young shell-like SNRs where a high speed of the order of 1000~km~s$^{-1}$ is often detected \citep[e.g.,][]{Hayato2010, Sato2017, Williams2018}.

The Galactic SNR W\,49B \citep{Westerhout1958, Mezger1967, Wynn-Williams1969} is most likely the youngest member of MM SNRs based on its small size of $\approx 4\arcmin\times3\arcmin$ or $\sim 10$~pc at the estimated distance of 11.3~kpc \citep{Brogan2001, Sano2021} and its high electron temperature of $\approx 1.5$~keV \citep{Ozawa2009}. W\,49B is host to an overionized/recombining plasma \citep{Ozawa2009, Yamaguchi2018} with one of the highest average charge states of Fe ions among all SNRs, i.e., the peak of the charge state distribution is between Fe\,{\scriptsize XXVI} and Fe\,{\scriptsize XXV} \citep{Yamaguchi2014}. W\,49B also exhibits hard X-rays originating from non-thermal bremsstrahlung of mildly energetic electrons with kinetic energies of the order of 10~keV \citep{Tanaka2018}. These features make this remnant arguably the most interesting example of MM SNRs. The distribution of the ejecta-dominated X-ray emission consists of a bar-like structure running through the center with flaring at its eastern and western ends in addition to a more circular and diffuse structure. The bar-like structure is more pronounced in Fe, while the diffuse, circular structure is pronounced in lighter elements such as Si and S \citep{Fujimoto1995, Keohane2007}. The longer wavelength counterparts create a barrel shape with coaxial rings in the near-infrared [Fe\,{\scriptsize II}] line and ear-like partial shells located at the eastern and western ends in the molecular hydrogen line, which are also bright in the radio continuum \citep{Moffett1994, Reach2006, Keohane2007}. The alignment of the bar-like Fe ejecta along the axis of the infrared coaxial rings was considered evidence for jets resulting from a bipolar explosion of a super-massive star \citep{Keohane2007, Lopez2013}. This interpretation, however, contradicts abundance measurements \citep[e.g.,][]{Hwang2000, Zhou2018, Sun2020, Sato2025, Sawada2025}, which suggest a Type Ia explosion or a Type II explosion of a relatively low-mass star. Therefore, the SN origin of this unique remnant is still unknown. 

A new observatory for X-ray astrophysics, the X-ray Imaging and Spectroscopy Mission (XRISM: \citealt{Tashiro2022}), successfully began observations in September 2023. XRISM carries two instruments: the Resolve high-resolution X-ray microcalorimeter spectrometer \citep{Ishisaki2022}, and the Xtend wide-band X-ray CCD imager \citep{Mori2022, Noda2025}. With the requirements of the spectral resolution of $\le 7$~eV at the full-width half maximum (FWHM) at 6~keV and the absolute energy-scale accuracy of $\le 2$~eV, Resolve enables us to resolve and identify many lines for the first time from a variety of highly ionized atoms in W\,49B, including, in many cases, fine-structure lines. These well resolved features can now be used to probe the line-of-sight (LOS) velocity of X-ray emitting objects such as SNR ejecta to the level of $\approx 100$~km~s$^{-1}$ in the Fe-K band. This is an improvement of over an order of magnitude compared to previous measurements. 

In this paper, we report the discovery of a systematic gradient in the LOS velocity along the Fe ejecta, which we claim is kinematic evidence for the bipolar flows of the ejecta. We also discuss the elemental dependence of the LOS velocity gradient and what that implies about the origin of W\,49B. The paper is structured as follows. The observations and data reduction will be described in \S\ref{sec:obs}. The spectral analysis and results from the Resolve spectrometer data will be presented in \S\ref{sec:ana}. The implications will be discussed in \S\ref{sec:disc}. Finally, the paper will be summarized in \S\ref{sec:summary}.

\section{Observations and data reduction} \label{sec:obs}

To cover the near entire $\approx 4\arcmin \times 3\arcmin$ remnant with the field of view (FOV) of $\approx 3\arcmin \times 3\arcmin$, XRISM's Resolve observed W\,49B with two aim points. The first, dubbed ``East'' (observation ID 300055010) started on April 23, 2024, while the second, dubbed ``West'' (observation ID 300056010) started on 30th April, 2024. Each observation lasted for about one week. 

Data analysis proceeded from cleaned events created using the pipeline-processing version 03.00.011.008 using the standard data-screening criteria\footnote{\url{https://heasarc.gsfc.nasa.gov/docs/xrism/analysis/abc_guide/XRISM_Data_Analysis.html}}. In this screening, time intervals with Earth occultations and passages of the South Atlantic Anomaly (SAA) were removed for both Resolve and Xtend data. In the case of Resolve, time intervals during the recycling of the adiabatic demagnetization refrigerator, those with the use of the onboard $^{55}$Fe calibration sources on the instrument filter wheel (FW) for the gain monitoring, and events for diagnostic purposes, e.g., baseline events and event-lost pseudo events, were also removed. In addition, non-X-ray events produced when cosmic rays interact with the Si frame around the detector pixels \citep[frame events:][]{Kilbourne2018} were identified and removed with their anomalously shaped detector pulses by using the event parameters \texttt{RISE\_TIME} and \texttt{DERIV\_MAX} according to \citet{Mochizuki2024}. The exact criteria on these parameters are described as the Resolve rise-time screening in the XRISM quick start guide v2.3\footnote{\url{https://heasarc.gsfc.nasa.gov/docs/xrism/analysis/quickstart/index.html}}. 

The pointing stability of the spacecraft during both observations were checked using the enhanced house-keeping (EHK) file included as part of the auxiliary files of the XRISM data products. A periodic excursion of the aim point associated with the orbital period of the spacecraft was discovered. In more than 90\% of the exposure times, the aim point exhibited an offset smaller than a few arcsec, while $\approx 30\arcsec$ offset was found in the remaining time. For each observation, the time intervals with the larger pointing offset were removed. The net exposure times for Resolve are 291.2~ks for East and 293.2~ks for West, while those for Xtend are 242.0~ks for East and 250.8~ks for West. Among the five Resolve X-ray event grades \citep{Ishisaki2018}, only the high-res events were used throughout this paper, whose fraction was higher than 93\% in the screened data for both observations.

The on-orbit time-dependent gain for each of the Resolve detector pixels was monitored and corrected using Mn K$\alpha$ from the $^{55}$Fe calibration sources on the FW using the standard XRISM/Resolve method described by \citet{Porter2024}. There is a known anomalous behavior for pixel 27 and it was excluded in the spectral analysis. For each pixel, the measured line centroid shift and its statistical uncertainty after the drift correction are $\le$ 0.05~eV at 5.9~keV during the fiducial intervals with the FW $^{55}$Fe sources. The Resolve detector also contains a calibration pixel which is part of the focal plane but located just outside of the instrument aperture and is illuminated continuously with a finely collimated $^{55}$Fe source. We use the calibration pixel to monitor the efficacy of the time dependent reconstruction of the energy scale during the main observation outside of the fiducial intervals. For East, the reconstruction error was 0.39~eV at 5.9~keV and for West 0.11~eV. We add this in quadrature with an estimate of the observation-independent energy scale uncertainty obtained by using additional on-board calibration sources \citep{Eckart2024} of 0.3~eV across the band 5.4--9.0~keV. This yields an energy scale uncertainty of 0.49~eV for East, and 0.32~eV for West. In addition, the per-pixel and composite-array energy resolution were measured as 4.0--5.5~eV and 4.5~eV FWHM, respectively, at 5.9~keV for high-resolution events using the $^{55}$Fe sources during the gain fiducial intervals\footnote{\url{https://heasarc.gsfc.nasa.gov/docs/xrism/analysis/gainreports/index.html}}. The resolution was stable during the observations as monitored continuously using the calibration pixel. The per-pixel energy resolution uncertainty for Resolve has been measured on the ground and in-flight and is energy dependent but corresponds to $< 0.15$~eV FWHM at 6~keV and $< 0.3$~eV at 10~keV.

\begin{figure*}[htbp]
\includegraphics[width=\textwidth]{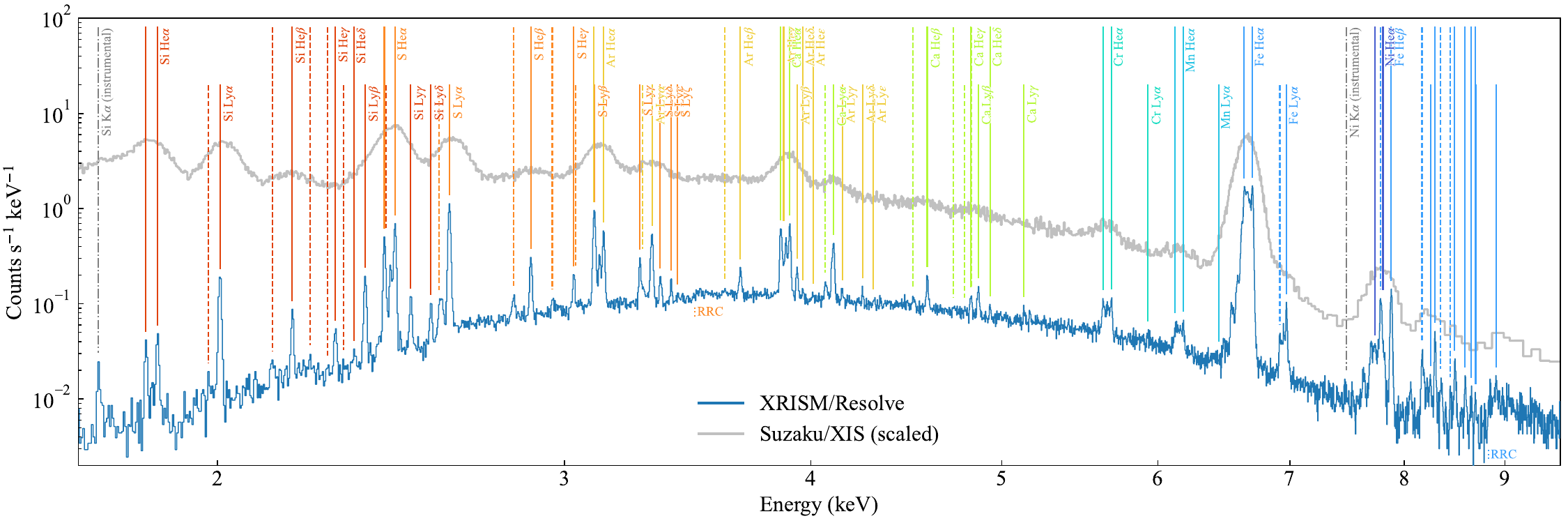}
\includegraphics[width=\textwidth]{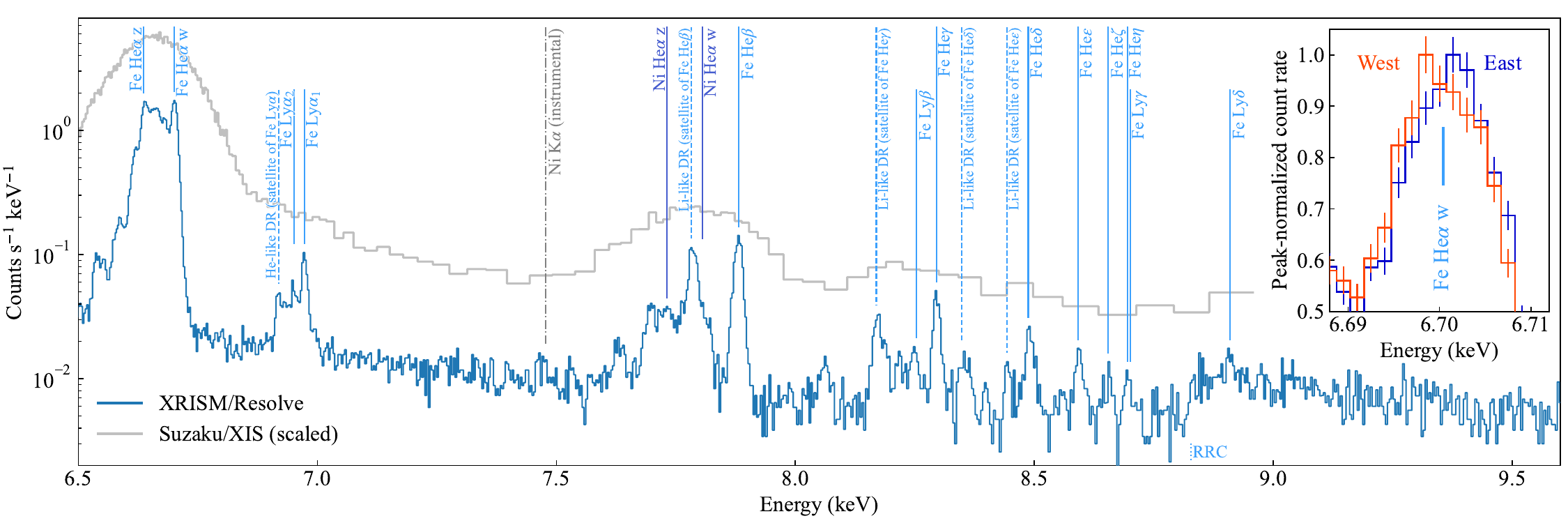}
\caption{The integrated spectra of W\,49B with Resolve (blue histogram). Only the high-resolution events are included. Emission features from ions of different elements are labeled: solid lines for Rydberg series lines (He$\alpha$, Ly$\alpha$, and higher-shell transition lines), dashed lines for dielectronic recombination (DR) satellite lines, and dotted lines for H-like S\,{\scriptsize XV} RRC and He-like Fe\,{\scriptsize XXV} RRC. Compared with the Resolve result is the spectrum with Suzaku X-ray Imaging Spectrometer (XIS: gray histogram) from \cite{Sawada2025}. The inset in the lower panel shows the close-up view of the Fe He$\alpha$ line plotted for the two observations separately.}
\label{fig:integspec}
\end{figure*}

\section{Analysis and results} \label{sec:ana}

\subsection{Integrated spectrum}

We first examined the Resolve full-array spectrum merged for the two observations, shown in Figure~\ref{fig:integspec}. In the top panel, emission lines from ions of Si, S, Ar, Ca, Cr, Mn, and Fe are clearly resolved, with possible underlying contribution from ions of other elements such as Ni. These spectral lines consist mainly of the strong He$\alpha$ complex consisting of the 2p$\rightarrow$1s resonance and intercombination lines, and the 2s$\rightarrow$1s forbidden line, known as lines w, x, y, and z, respectively. The hydrogen-like 2p$\rightarrow$1s Ly$\alpha$ doublets, and higher-shell Rydberg transitions ($n$p$\rightarrow$1s with $n\ge3$, denoted as He$\beta$, He$\gamma$, and so on, for He-like ions for instance), and their dielectronic recombination (DR) satellite lines are also observed. Radiative recombination continuum (RRC) edges are clearly detected for Fe\,{\scriptsize XXV} at 8.8~keV and S\,{\scriptsize XVI} at 3.5~keV. As evident in the bottom panel, the $n\ge3$ Rydberg lines from He-like ions are detected at least up to $n=8$ (He$\eta$) in the case of Fe\,{\scriptsize XXV}, and these are accompanied with DR satellites of comparable intensities. Together with the strong RRC, high-$n$ excitation lines and strong DR satellites are characteristics of a recombining plasma \citep{Kaastra2008, Sawada2025}, which is well showcased in the Resolve spectrum.

The inset in the lower panel of Figure~\ref{fig:integspec} is a close-up view of the peak-normalized Fe He$\alpha$ resonance (w) lines from the two observation fields. Despite a large overlap of the observation fields at the bright center of the remnant, the resonance line peaks are slightly but significantly shifted and skewed in the opposite directions, redward for West and blueward for East. This indicates a systematic LOS velocity variation across the Fe ejecta. The peak energy difference is about two energy bins in the plot, or 3~eV, corresponding to a possible LOS velocity separation of $\approx$ 130~km~s$^{-1}$. 

\subsection{Pixel-to-pixel Fe He$\alpha$ spectra} \label{sec:pix2pix}

\begin{figure*}[htbp]
\begin{center}
\includegraphics[width=\columnwidth]{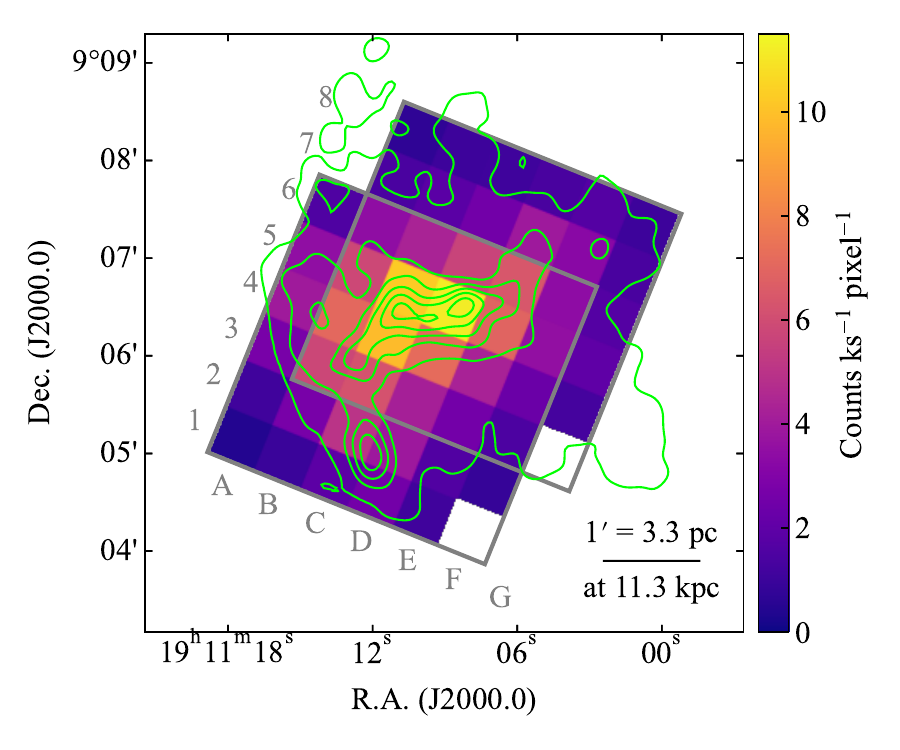}
\includegraphics[width=\columnwidth]{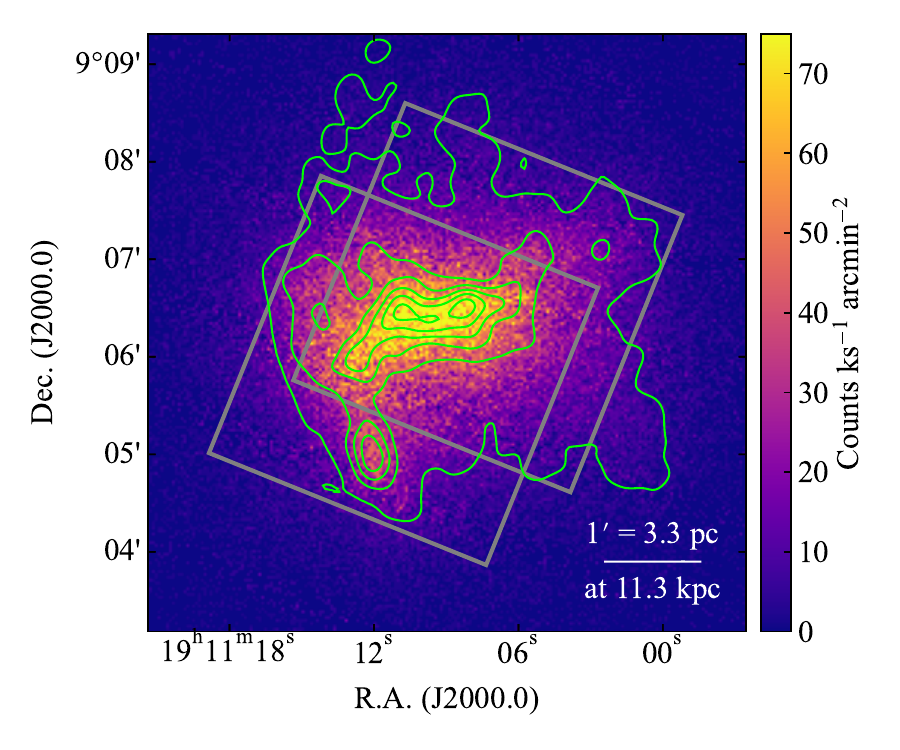}
\caption{The XRISM Fe He$\alpha$ band images of W\,49B with Resolve (left, in 6.53--6.80~keV) and Xtend (right, in 6.4--6.9~keV). The two large squares show the East and West Resolve FOVs. The contours are generated from Chandra's ACIS 6.3--7.0~keV. The Resolve image is affected by both the blurring due to the mirrors' PSF and the relatively sparse $6\times 6$ detector array. In the Xtend, the spatial resolution is entirely determined by the mirror PSF.}
\label{fig:images}
\end{center}
\end{figure*}

\begin{figure*}[htbp]
\begin{center}
\includegraphics[width=0.32\textwidth]{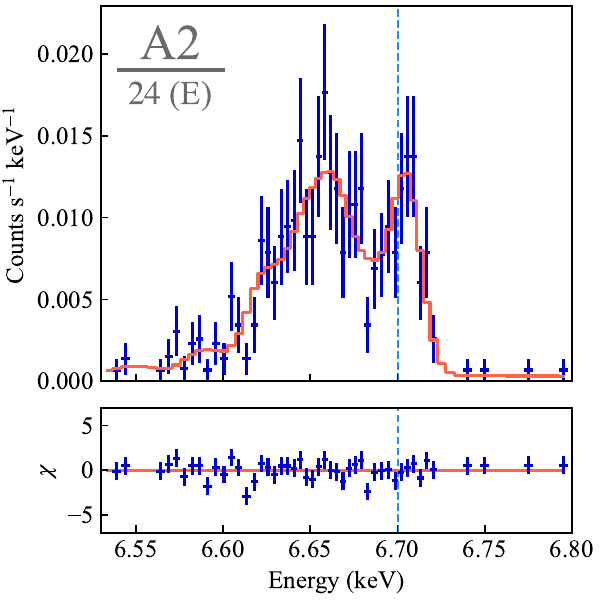}
\includegraphics[width=0.32\textwidth]{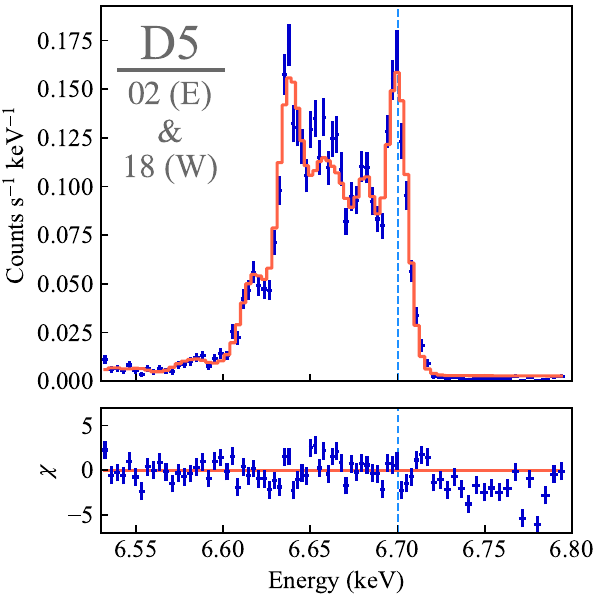}
\includegraphics[width=0.32\textwidth]{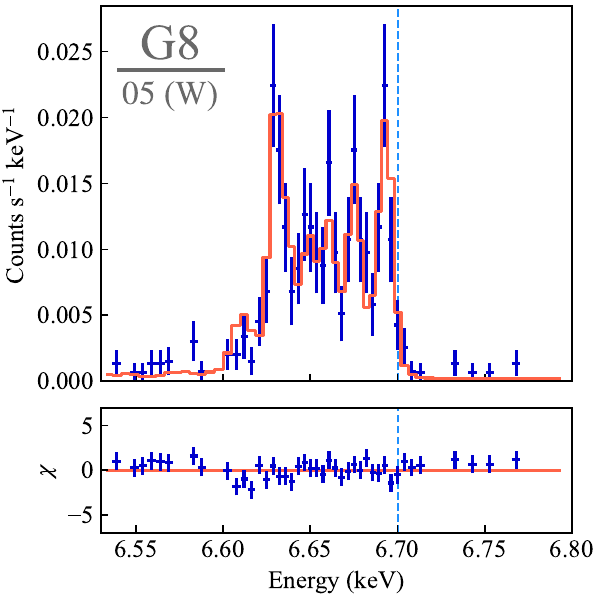}
\caption{Examples of pixel-by-pixel Fe He$\alpha$ spectral fits (top panels) from the A2, D5, and G8 regions (Figure~\ref{fig:images}) with residuals shown in the form of (data $-$ model)$/$(1$\sigma$ error) (bottom panels). The dashed vertical lines show the rest-frame energy of Fe He$\alpha$ resonance line (w), which corresponds to the highest-energy peak in the data shown. The numbers below the region name show the detector pixels used, E for East and W for West.}
\label{fig:pixelspec}
\end{center}
\end{figure*}

To examine the possible LOS velocity variation, we next analyzed pixel-to-pixel Resolve spectra. The projected locations of Resolve FOVs and pixel positions are shown with the Resolve and Xtend images of Fe He$\alpha$ in Figure~\ref{fig:images}. The contours compared to these are from the Chandra Advanced CCD Imaging Spectrometer (ACIS) 6.3--7.0~keV map using $4 \times 4$ binning and smoothed with a 4$\arcsec$ Gaussian kernel. The image is from the Chandra Supernova Remnant Catalog\footnote{\url{https://hea-www.harvard.edu/ChandraSNR/G043.3-00.2/}}.  

The two fields have significant overlap at the center of the remnant. Thus, in regions of the sky covered by both observations, we merged single-pixel spectra from the two fields into one spectrum. The alignment error of the pixel boundaries between the two fields was only $1.2\arcsec$. The redistribution matrix file (RMF) was generated for each pixel using the \texttt{rslmkrmf} tool. Throughout this paper, the large-type RMF was used, hence all the modeled detector-relevant matrix components except for the electron-loss continuum are included. This choice is appropriate for this study because we concentrate on the narrow-band spectral fits where the electron-loss continuum component does not affect our results. For regions covered by the both observations, an exposure-time-weighted average of the RMFs for each pixel pair was taken. For the auxiliary response file (ARF), we used the one simulated for on-axis point sources observed with the full array of Resolve. This would significantly overestimate the effective area when analyzing a single-pixel spectrum, however, it is appropriate here because the absolute normalization of the spectrum does not correlate with the LOS velocity or broadening. In other words, a possible systematic deviation in the relative effective area (or its energy dependence) arising from the assumption of a point-like source distribution instead of an extended source distribution is too small to alter fitting results in the narrow-band analysis.

The Fe He$\alpha$ spectrum from each Resolve pixel was fitted in the 6.53--6.80~keV range with the recombining plasma model \texttt{brnei} in the spectral fitting package Xspec \citep{Arnaud1996} with the atomic code AtomDB version 3.0.9 \citep{Smith2001, Foster2012, Foster2020}. We note that switching the atomic code to the recently released AtomDB versions 3.1.0--3.1.2 or using an alternative spectral fitting package and atomic code SPEX version 3.08.01 \citep{Kaastra1996, Kaastra2024} with its non-equilibrium ionization plasma model \texttt{neij} does not affect the results, discussions, or conclusions presented in this paper. The model parameters are the emission measure ($EM$) as a normalization of a spectrum, the electron temperature ($kT_{\rm e}$), the initial temperature ($kT_{\rm init}$) and recombination timescale ($\tau_{\rm rec}$) to give the over-ionized charge-state distribution, the abundance ($Z$), the redshift ($z$), and the velocity dispersion ($\sigma_V$) describing the line broadening. Note that the thermal broadening is not modeled because the ion temperature is unknown in an SNR where the electron-ion temperature equilibrium is not necessarily reached. The abundance (effectively the Fe abundance) was fixed at 5 $\times$ the solar values \citep{Lodders2009} because the continuum level cannot be determined for some of the pixels with relatively low statistics. Also fixed during the fit was the initial temperature set at 4~keV \citep[e.g.,][]{Yamaguchi2018}.

\begin{figure*}[htbp]
\begin{center}
\includegraphics[width=\textwidth]{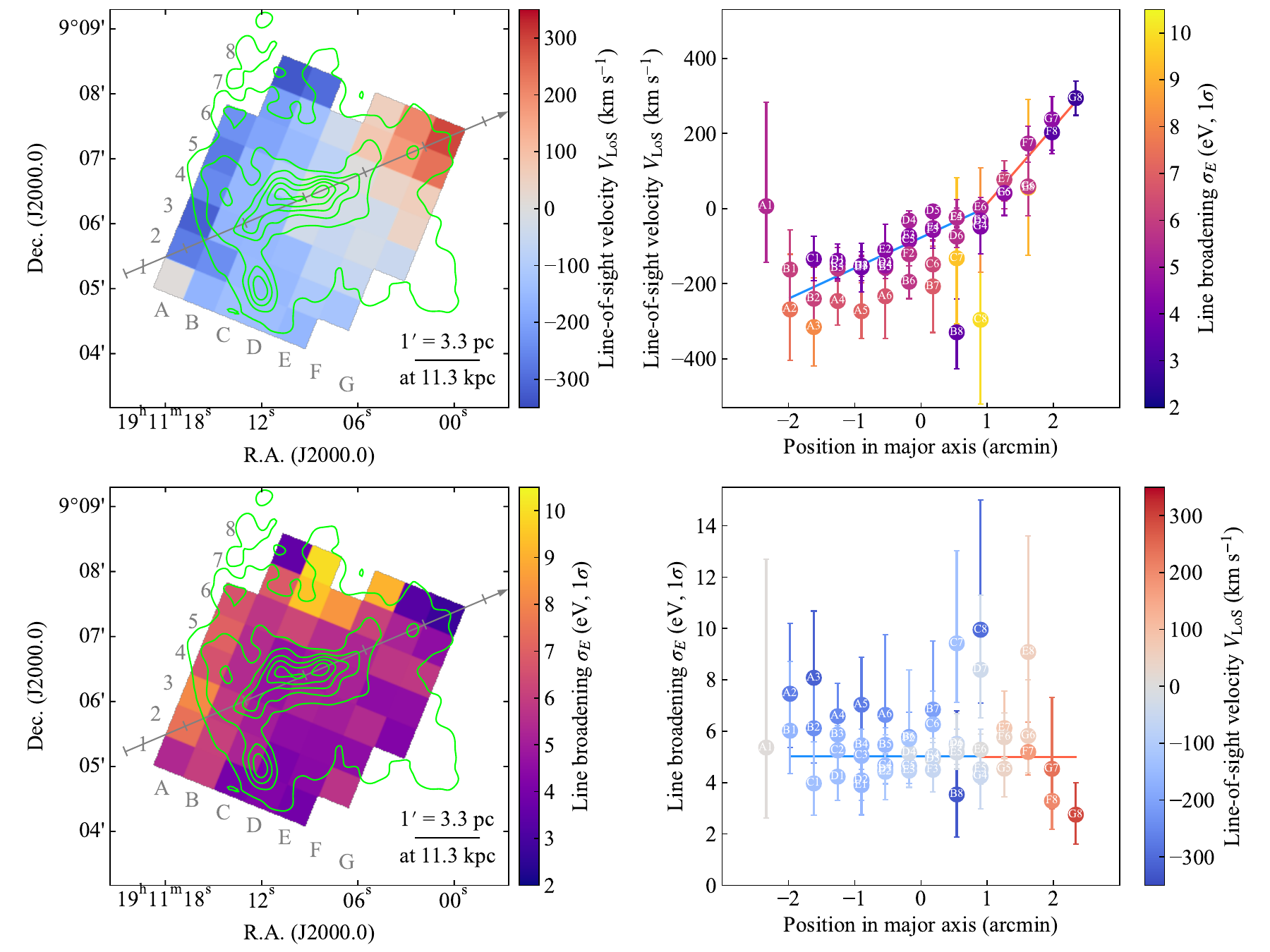}
\caption{The LOS velocity (top) and line broadening (bottom) measured at Fe He$\alpha$. The left panels show pixel maps with the ticked arrows tracing the approximate major axis of the Fe ejecta, while the right panels show values with the 90\% statistical errors as a function of the position along the major axis.}
\label{fig:pixana}
\end{center}
\end{figure*}

Three examples of the spectral fits are shown in Figure~\ref{fig:pixelspec}. Each of these was taken from one Resolve detector pixel along the bright Fe ejecta bar. The dashed vertical lines in the spectral plots indicate the rest-frame energy of the Fe He$\alpha$-w line. The observed peaks show significant shifts by 5--6~eV. A blueshift is detected in a southeastern region (A2), while it is rather a redshift in a northwestern region (G8). The spectrum of the pixel located at the middle of the remnant (D5) appears to have almost no energy shift. It is also worth noting that the middle spectrum does not show a clear sign of double-peak structure, which would be expected toward the center of an expanding shell. Note that the region A2 near the south-eastern corner appears to be located outside the sharp limb in the Chandra 6.3--7.0~keV image, but actually has significant counts from the limb due to the point-spread function (PSF) of XRISM's X-ray mirrors with a half-power diameter of $1.3\arcmin$ \citep{Hayashi2024}. 

To have a full picture of the velocity structure, we derived pixel maps of the LOS velocity (obtained as $V_{\rm LOS} = cz$, where $c$ is the speed of light) and line broadening ($\sigma_E = \sigma_V E / c$, where $E$ is the rest-frame energy of the resonance line of Fe He$\alpha$, 6700.42~eV) as in the left panels of Figure~\ref{fig:pixana}. Here, the line broadening is expressed in the energy space rather than in the velocity space. This is because, unlike the line centroid shift whose only possible astrophysical origin for a Galactic source is the LOS velocity, the broadening can also be contributed by non-kinematic factors such as thermal broadening. The LOS velocity shows a systematic gradient from blueshift (negative $V_{\rm LOS}$) in the east to redshift (positive $V_{\rm LOS}$) in the west. The map also shows that pixels with the larger blueshift values were found around the northern corner (such as B7 and B8 in Figure~\ref{fig:images}) and northeastern edge (such as A2--A6 and B2) of the Resolve FOVs, indicating a local structure on top of the overall east-west gradient. On the other hand, the line broadening does not show east-west gradient and seems to have a nearly constant value except a few edge regions to the north. Those with the largest broadening (C7--C8, D7, and E8) lie between the boundary of the large blueshift and redshift regions, indicating that the large broadening may be the result of the relatively large PSF which causes a spatial mixing of the spectral components, i.e., spatial-spectral mixing of different velocity components. 

The LOS velocities and line broadenings are plotted in the right panels of Figure~\ref{fig:pixana}, each as a function of the position in the major axis of the Fe ejecta. The LOS velocity shows a nearly monotonic change along the major axis of the ejecta bar, confirming the east-west gradient.  It appears that the gradient is steeper in the redshifted part in the west than in the blueshifted part in the east. By fitting a linear function to the $V_{\rm LOS}$ distribution in the top right panel of Figure~\ref{fig:pixana}, we derived the slope and intercept for redshifted and blueshifted groups as shown with solid lines. In this evaluation, we ignored A1, B8, and C8 as these are outliers. The outliers were identified by comparing $V_{\rm LOS}$ of each pixel to adjacent pixels along the major axis. Thus evaluated local deviations in $V_{\rm LOS}$ were 220--250~km~s$^{-1}$ for these three pixels, while those for the remaining pixels were 130~km~s$^{-1}$ at largest and distributed in $\approx 60\pm40$~km~s$^{-1}$ (standard deviation). We note that exclusion of the outliers in this analysis does not significantly impact the results much because of the large statistical errors. 
The slope for the redshifted group was $62 \pm 9$~km~s$^{-1}$~pc$^{-1}$, while that for the blueshifted group was $25 \pm 3$~km~s$^{-1}$~pc$^{-1}$, for the distance of 11.3~kpc. These differ significantly from each other by a factor of two. The slope derived including both groups was $32 \pm 2$~km~s$^{-1}$~pc$^{-1}$. The overall line broadening distribution, in contrast to the LOS velocity, shows no significant gradient with the best-fit slope of $0.035 \pm 0.045$~eV~pc$^{-1}$. A search for the east-west difference was conducted also for the line broadening, but in this case using a constant model to compare the average broadenings. The two groups showed an identical average broadening of $\sigma_E = 5.0$~eV as shown with solid lines in the bottom right of Figure~\ref{fig:pixana}. This corresponds to $\sigma_V = 220$~km~s$^{-1}$ if it is fully attributed to the velocity dispersion, or alternatively, an ion temperature of $\approx 30$~keV for Fe if it is fully attributed to the thermal broadening. We note that, the redshifted group showed an indication of a local gradient of $-0.46 \pm 0.23$~eV~pc$^{-1}$, which is a 3.3$\sigma$ deviation from a constant broadening case. We emphasize that, with the resolution of 4.5~eV at FWHM (or 1.9~eV at 1$\sigma$), the major structures of the Fe He$\alpha$ complex with $\sigma_E = 5$~eV are well resolved as in the middle panel of Figure~\ref{fig:pixelspec}, and therefore, the lack of a gradient in the overall line broadening  distribution is not because of insufficient sensitivity to detect enhanced broadening but because of the absence of a systematic gradient in the intrinsic broadening except for a possible local trend in the northwest. 

In this pixel-by-pixel spectral analysis, the best-fit values of the thermal parameters were also obtained. Significant variations were found in the electron temperature ranging in $kT_{\rm e}=1.1\textrm{--}1.7$~keV and in the recombination timescale in $\tau_{\rm rec}=(1\textrm{--}6)\times10^{11}$~cm$^{-3}$~s. These are in good agreement with the previous studies \citep[e.g.,][]{Yamaguchi2018}. The detailed results on the thermal parameters and their spatial distributions will be presented elsewhere.

\subsection{LOS velocity of various elements} \label{sec:elemdep}

\begin{figure*}[htbp]
\begin{center}
\includegraphics[width=\textwidth]{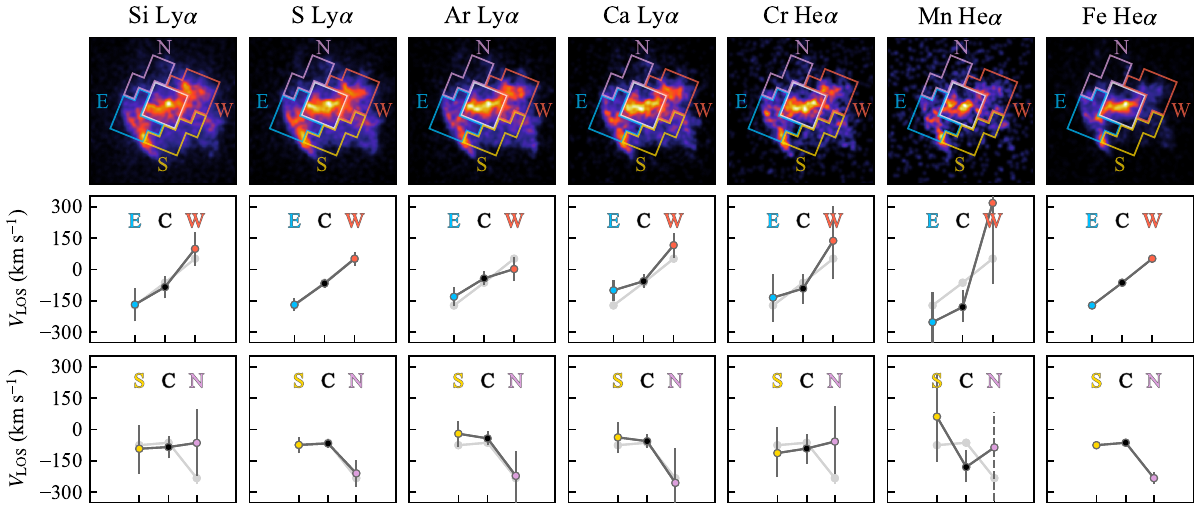}
\caption{The results of the line-of-sight velocity $V_{\rm LOS}$ for various elements. Top panels show the five groups of the Resolve detector pixels with  Chandra images (Si Ly$\alpha$ in 1.92--2.12~keV, S Ly$\alpha$ in 2.58--2.78~keV, Ar Ly$\alpha$ in 3.26--3.48~keV, Ca Ly$\alpha$ in 4.05--4.29~keV, Cr He$\alpha$ in 5.45--5.90~keV, Mn He$\alpha$ in 5.90--6.30~keV, and Fe He$\alpha$ in 6.30--7.00~keV). Middle and bottom panels show the $V_{\rm LOS}$ variations along the east-west and north-south directions, respectively. The darker error bar plots are the results of the entitled energy bands, while the lighter-colored error bars behind them are those for Fe He$\alpha$ for comparison. The errors are quoted at the 90\% confidence level except for the one with a dashed bar, which is at the 68\% level. See text for the details.}
\label{fig:elemdep}
\end{center}
\end{figure*}

The spatial distribution of ejecta is systematically different between elements \citep{Fujimoto1995,Keohane2007}. It is natural for one to expect that such a difference exists also in the velocity distribution. Because the statistics of other lines are limited compared to Fe He$\alpha$, we grouped Resolve detector pixels to compose five regions covering the center (C), north (N), south (S), east (E), and west (W) of the remnant, and analyzed their spectra in several narrow energy bands, encompassing Ly$\alpha$ lines of Si, S, Ar, and Ca as well as He$\alpha$ of Cr, Mn, and Fe. Note that, among the 49 pixel regions analyzed in \S\ref{sec:pix2pix}, the five pixel regions in the row ``1'' in Figure~\ref{fig:pixana} were excluded in this analysis to make the locations of the E and W regions with respective to the C region more symmetric. The preference of H-like Ly$\alpha$ to He$\alpha$ for intermediate-mass elements (Si, S, Ar, and Ca) is because Ly$\alpha$ is comparably bright to He$\alpha$ and it is relatively free from mixing with lines of other elements; e.g., He$\alpha$ of S and Ar are overlapping with a DR satellite of Si and Ly$\beta$ of S, respectively, as seen in Figure~\ref{fig:integspec}. 
The preference of He$\alpha$ to Ly$\alpha$ for Fe-group elements (Cr, Mn, and Fe) is because Ly$\alpha$ is much fainter. We thus chose the energy bands to be 1.99--2.03~keV for Si, 2.60--2.64~keV for S, 3.30--3.34~keV for Ar, 4.08--4.14~keV for Ca, 5.55--5.75~keV for Cr, 6.05--6.25~keV for Mn, and 6.53--6.80~keV for Fe. 

We first performed a baseline fit for each region in 6.53--7.02~keV covering both Fe He$\alpha$ and Fe Ly$\alpha$ to determine the thermal parameters as well as the emission measure, and then analyzed individual narrow-band spectra to determine the LOS velocities and line broadenings. The reason why the continuum was not adopted in the determination of the thermal parameters in the baseline fit is mainly because it would be complicated by the presence of the non-X-ray background, while we confirmed that the changes in the thermal parameters by including the continuum in the 5.0--7.5~keV band but excluding the Cr and Mn He$\alpha$ lines are not significant ($\approx 1\sigma$ at largest) even without properly modeling the non-X-ray background. In both analysis steps, we used \texttt{bvvrnei} in place of \texttt{brnei} to allow elemental abundance to have separate values depending on elements. We also employed \texttt{tbabs} to reproduce the interstellar absorption at low energies assuming the atomic hydrogen equivalent column density of $5\times 10^{22}$~cm$^{-2}$ \citep{Keohane2007}. In the baseline fit, we set the initial temperature at 4~keV as in the pixel-by-pixel analysis (\S\ref{sec:pix2pix}), and made the iron abundance free in addition to the five free parameters ($EM$, $kT_{\rm e}$, $\tau_{\rm rec}$, $z$, and $\sigma_V$). The best-fit thermal parameters in the baseline fit were all within the ranges of the variations found in the pixel-by-pixel analysis of Fe He$\alpha$ (\S\ref{sec:pix2pix}). 
In the individual fits, we fixed the three thermal parameters ($kT_{\rm e}$, $kT_{\rm init}$, and $\tau_{\rm rec}$) as well as $EM$ to those in the baseline fit, while making the other parameters free ($z$, $\sigma_V$, and the abundance of the element of interest). We note that modifying the thermal parameters within ranges of possible spatial or energy-band dependence does not affect the measured LOS velocities significantly. For instance, if we take the Si Ly$\alpha$ fit of the C region as an example, we get only a small change of 10~km~s$^{-1}$ in the LOS velocity corresponding to the 0.3$\sigma$ significance level, even with a drastic change of $kT_{\rm e}$ from 1.5~keV with the baseline fit to 0.5~keV suggested as a low-temperature component partly contributing to the Si-K band based on wide-band CCD spectroscopy \citep[e.g.,][]{Holland-Ashford2020}.

The LOS velocity $V_{\rm LOS}$ measured with the seven elements are shown in Figure~\ref{fig:elemdep}. An immediate result is that the same east-west gradient as found in the pixel-by-pixel analysis of Fe He$\alpha$ is confirmed with all the elements (middle panels). There is a hint of a small variation between elements, e.g., Mn may have a larger blueshift in the regions E and C and a larger redshift in the region W compared to Fe, but these are statistically marginal. In the north-south direction (bottom panels), the change is not as monotonic as in the east-west direction. The regions C and S both show a mild blueshift with  $V_{\rm LOS} \approx -70$~km~s$^{-1}$, while the region N shows a larger blueshift of $V_{\rm LOS} \approx -230$~km~s$^{-1}$ at Fe He$\alpha$, with which the other elements show consistent values within the statistical errors. The blueshift in the region N is even greater than that in the region E. This trend is consistent with the pixel-by-pixel analysis results (Figure~\ref{fig:pixana} top left) where we found pixels with the largest blueshift at the northeastern edges. We note that, for Mn in the region N, the 68\% error is shown because the significance of the emission is too low to determine the 90\% error.

\section{Discussion} \label{sec:disc}

\subsection{Kinematics of ejecta in W\,49B} \label{sec:kinem}

\begin{figure*}[htbp]
\begin{center}
\includegraphics[width=\textwidth]{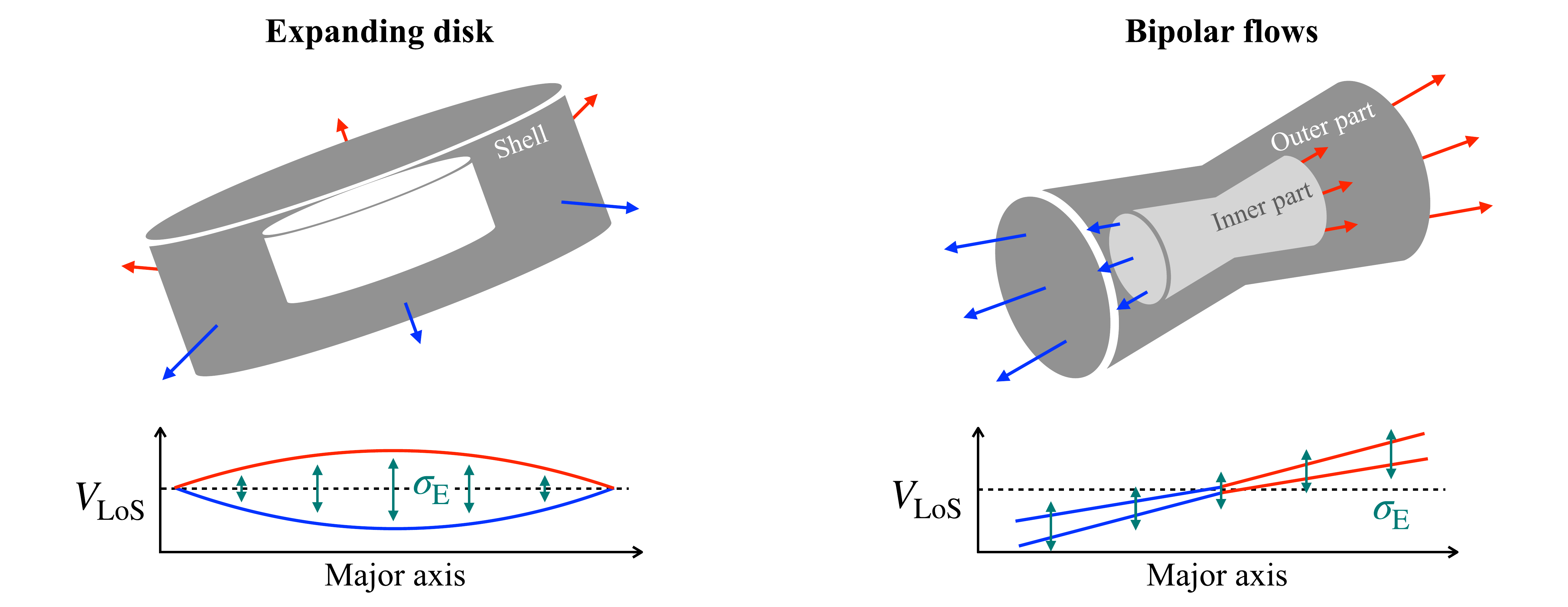}
\caption{A schematic of the expected LoS velocity and broadening distributions for representative two cases of the ejecta kinematics: equatorially expanding disk (left) and bipolar flows (right). The data and accompanying analysis presented here reject the expanding disk model.}
\label{fig:schematic}
\end{center}
\end{figure*}

We first discuss the kinematics of the ejecta in W\,49B based on the LOS velocity and broadening measurements. The elongated bar-like morphology of the Fe ejecta is a unique feature of this remnant, which gives rise to various interpretations. In terms of the kinematics, these are broadly divided into two models. In one model, the ejecta are assumed to have bipolar flows along the major axis and the jet-like morphology is due to this intrinsic asphericity \citep{Keohane2007, Lopez2013, Gonzalez-Casanova2014}. In the other model, the morphology is rather attributed to the enhanced reverse shock due to an aspherical circumstellar matter (CSM) or molecular cloud \citep{Miceli2006, Miceli2008, Shimizu2012, Zhou2018}, which may be supported by the existence of shock-excited clouds \citep[e.g.,][]{Zhou2022}. In this case, the motion of the ejecta should be dominated by a spherical or equatorial expansion of disk-like shocked ejecta. These two kinematic models are summarized in Figure~\ref{fig:schematic}. 

We argue that the expanding disk model (Figure~\ref{fig:schematic} left) is rejected by the presented measurements with XRISM. In this model, both the redshifted and blueshifted components should be seen along the ejecta bar, with the maximum LOS velocities at the center and the minimum at both ends. This expectation clearly contradicts the measured LOS velocity distribution (Figure~\ref{fig:pixana} top right). The spectrum observed near the center of the remnant does not show a line splitting or double-peaked profile originating from the redshifted and blueshifted components (Figure~\ref{fig:pixelspec} middle). One may argue that the line profile may depend on the radial distribution of the emission measure and velocity; i.e., if the density is higher toward the inner radii as suggested by the centrally peaked X-ray morphology, and the expansion is homologous, then the emission lines may still have a single peaked profile. In such a case, large LOS velocity components corresponding to outer part of the expanding disk contributes to the broadening of the lines. Then, because the magnitude of the overall LOS velocity decreases toward eastern and western ends of the ejecta, we would observe a monotonic decrease in the broadening outward. However, our measurement showed that the line broadening at Fe He$\alpha$ is nearly uniform (Figure~\ref{fig:pixana} bottom right). 

The bipolar flows, on the other hand, naturally explain the observed LOS velocity distribution (Figure~\ref{fig:schematic} right). If bipolar flows are slightly angled as in the schematic, the emission from the approaching side to us shifts blueward and that from the receding side from us shifts redward, reproducing the systematic difference in $V_{\rm LOS}$ between the eastern and western parts (Figure~\ref{fig:pixana} top right). The smooth gradient in the LOS velocity can be explained if the bipolar flows are homologous, i.e., $V_r \sim r / t_{\rm age}$, where $r$ is the radial (projected) distance from the SNR center, $V_r$ is the radial (or projected) velocity, and $t_{\rm age}$ is the SNR age. The inclination angle of outflows with respect to the LOS direction is $\theta \approx V_{\rm LOS} / V_r = V_{\rm LOS}~t_{\rm age}/ r$, which is
\begin{equation}
\theta~(\arcdeg) \approx 6 \left(\frac{V_{\rm LOS}}{\rm 100~km~s^{-1}}\right) \left(\frac{r}{\rm 1~pc}\right)^{-1} \left(\frac{t_{\rm age}}{\rm 10^3~yr}\right).
\end{equation}

\noindent
For the estimated age of $t_{\rm age} \sim 5$~kyr \citep{Hwang2000, Zhou2018} and the LOS velocity of 300~km~s$^{-1}$ at the outermost part at $r = 6.6$~pc (semi-major axis), we get $\theta \approx 14\arcdeg$. If we derive the inclination angle separately for the blueshifted and redshifted regions using the best-fit slopes (top right of Figure~\ref{fig:pixana}), $V_{\rm LOS}/r = 25$~km~s$^{-1}$~pc$^{-1}$ and $62$~km~s$^{-1}$~pc$^{-1}$, respectively (\S\ref{sec:pix2pix}), it is slightly shallower ($\theta \approx 8\arcdeg$) for the east and deeper ($\theta \approx 19\arcdeg$) for the west. 

We note that the smooth gradient in the LOS velocity is not due to the spatial spectral mixing (SSM), by which spectral components arising from spatially different locations are mixed due to the PSF of the mirrors. If the actual LOS velocity distribution consists of two distinct groups of blueshift and redshift without having the intermediate group with almost no LOS velocity, then the nearly zero energy shift observed toward the center is due to a mixing of the blueshifted and redshifted components. As already discussed for the expanding disk model, with such a mixing, we expect an enhancement of the broadening toward the center, which is not the case. Therefore, the smooth gradient cannot be explained by two distinct velocity groups affected by the SSM effect and should be an intrinsic property. This view is consistent with a nearly uniform broadening observed along the ejecta bar because the magnitude of the broadening due to the SSM effect is determined by the gradient of the LOS velocity, which is nearly constant. However, the velocity dispersion, obtained if the broadening (mostly 3--7~eV) is fully attributed to kinematic origins, is $\approx 220 \pm 90$~km~s$^{-1}$, which is larger than the scale of the LOS velocity variation (100--200~km~s$^{-1}$) within the half-power diameter of the PSF ($\sim 1\arcmin$), suggesting significant contribution from other origins such as thermal motion of ions. Distinguishing the origin of the broadening requires further analysis considering the SSM effect and will be reported elsewhere. 

\subsection{Implications to origin of W\,49B}

We next discuss possible origins of the W\,49B. As concluded in \S\ref{sec:kinem}, the bipolar velocity distribution cannot be reproduced with spherically or equatorially expanding, disk-like ejecta. This fact actually rejects most of the scenarios previously proposed. 

\subsubsection{Scenarios predicting equatorially expanding ejecta} \label{sec:equatorial}

\citet{Zhou2018} proposed an asymmetric Type Ia model to simultaneously explain the abundances pattern and spatial distribution of heavy elements. In their model, the bar-like morphology was attributed rather to the anisotropy in the ambient mass density, which was assumed to be higher in the equatorial, east-west direction. This resulted in earlier and stronger reverse-shock formation in the east-west direction, compared to the north-south direction. This is similar to the model considered by \citet{Miceli2006, Miceli2008}. In such a case, however, the expected kinematics should be an equatorially expanding disk (Figure~\ref{fig:schematic} left), which is inconsistent with the observed $V_{\rm LOS}$ distribution with XRISM (\S\ref{sec:kinem}). 

A spherical core-collapse (CC) origin would be the next candidate \citep{Hwang2000, Sun2020, Sawada2025} in terms of the elemental abundances. \citet{Shimizu2012} reproduced the bar-like X-ray morphology as well as the recombining plasma in their numerical simulations of a spherical CC explosion inside an axially symmetric, disk-like CSM. In this scenario, the structure of the ejecta is an equatorially expanding disk (Figure~\ref{fig:schematic} left), which is again inconsistent with the observed $V_{\rm LOS}$ distribution with XRISM (\S\ref{sec:kinem}). 

Recently proposed common-envelope jets SN (CEJSN) scenario \citep{Grichener2023} considers a thermonuclear outburst in a massive accretion disk around a neutron star, formed by tidal disruption of the core of a companion red supergiant. This model explains both the Type Ia-like nucleosynthesis and the bar-like ejecta distribution. However, in this model, the ejecta originate from the accretion disk rather than the bipolar jets launched in the direction perpendicular to it. The expected LOS velocity distribution of the ejecta, therefore, should be closer to that of an equatorially expanding disk (Figure~\ref{fig:schematic} left), which is clearly inconsistent with the XRISM result. 

There are some scenarios that do not specify a particular explosion type or a progenitor mass. One of such is a model by \citet{Zhang2019}, where a SNR was assumed to evolve in an ambient medium filled by many clumps of molecular clouds. While the kinematics of the ejecta may be altered by collisions with the dense clumps, it is unlikely that the cloud interaction acts to align the ejecta velocity to make bipolar flows. Therefore, even if the interaction plays a significant role to reproduce other observational characteristics of W\,49B, there must be another mechanism responsible for bipolar flows.

\subsubsection{Bipolar SN explosion}

A certain type of massive star is expected to undergo jet-driven bipolar explosions \citep[e.g.,][]{Maeda2003}, which has been proposed to be the cause of the bar-like morphology \citep{Keohane2007, Lopez2013}. The XRISM discovery of the bipolar kinematics (Figure~\ref{fig:schematic} right) can easily be associated with this scenario. We note that, unlike the previously proposal by \citet{Lopez2013}, a bipolar CC SN does not necessarily mean a hypernova origin. For instance, evidence for bipolar explosions in Type IIP SNe have been reported recently \citep{Nagao2024}.

A numerical simulation showed that, in this model, the bar-like structure is more prominent in Fe, while the more circular component is brighter in lighter elements such as Si and S \citep{Gonzalez-Casanova2014}, which matches well with the observed morphologies in various energy bands. This may seem to contradict the common LOS velocity structures between elements (\S\ref{sec:elemdep}). We argue that this is not necessarily the case. For instance, in the simulations by \citet{Gonzalez-Casanova2014}, the more circular morphology of the lighter elements such as Si and S was due to their farther distributions from the jet axis rather than the lack of the bipolarity. These elements actually share the same bipolarity as Fe, and therefore, we can expect similar LOS velocity distributions, especially with the spatial resolution of Resolve. We note that \citet{Gonzalez-Casanova2014} simulated the evolution only up to 700~yr based on an old age estimate by \cite{Pye1984}, which relied on the existence of synchrotron X-ray emission and is therefore no longer valid. The reproducibility of the bar-like structure at the age of W\,49B needs to be confirmed with a longer simulation time. 

One may dispute the jet-driven bipolar explosion model because the jet base should be displaced to make a clear ``gap'' at the center of the remnant, which appears in contrast to the centrally peaked X-ray distribution in the Xtend image (Figure~\ref{fig:images} right). However, in the finer-resolution Chandra image (contours in Figure~\ref{fig:images}), two bright spots near the center are evident, one at the pixel region D5 and the other between the pixel regions C4 and C5. These spots are connected with fainter emission filling the ``gap''. This emission can be explained by a projection effect as seen in the simulated X-ray maps by \cite{Gonzalez-Casanova2014}. The jet base has extended ring-like structures at outer radii of the jet axis. These structures of the two sides are superposed along the LOS at the center to produce X-rays from the ``gap''.
The separation of the bright spots is $\approx 0.6\arcmin$ or $\approx 2$~pc at 11.3~kpc, indicating the average velocity of the jet base of $\approx 190$~km~s$^{-1}$. Assuming an inclination of $14\arcdeg$ (\S\ref{sec:kinem}), the separation in the LOS velocity is expected to be $\approx 90$~km~s$^{-1}$, which is consistent with the observed separation between these pixel regions (Figure~\ref{fig:pixana}). This also produces a sufficiently small line-centroid splitting ($\lesssim 2$~eV) not to cause significant distortion or broadening in the observed spectrum at the center. Assuming that we interpret the two bright spots as the jet base, the jet center has a LOS velocity of $\approx -50$~km~s$^{-1}$, which is consistent with the fact that a larger fraction of the ejecta shows blueshift (Figure~\ref{fig:pixana}). The asymmetry may be caused by momentum taken away by the undiscovered central compact object, suggesting its runaway direction toward west. We note that the X-ray distribution in the simulations by \cite{Gonzalez-Casanova2014} still has some discrepancies from the observed one such as the jet tip being brighter than the jet base as opposed to the observation. 

Major difficulties of the bipolar SN scenario with a massive progenitor rather exist in explaining other aspects of this SNR, mainly the elemental abundances, such as the high Mn and low Ti abundances \citep{Zhou2018, Sato2025}. A recent spectroscopic study with Suzaku showed that the abundance ratio measurements of the Fe-group elements, especially Ni/Fe, depended on the choice of the atomic codes \citep{Sawada2025}. Together with other systematics originating from the moderate energy resolution of X-ray CCDs, it is possible that the previously reported results are biased. High-resolution spectroscopy is imperative to get decisive results. The feasibility of the massive progenitor bipolar explosion scenario will be revisited in a separate paper with updated abundance measurements with XRISM/Resolve.

Bipolarity is not necessarily a unique feature to CC SNe but could also be achieved by highly asymmetric Ia SNe. One possibility is gravitationally confined detonation model \citep{Plewa2004, Townsley2007, Jordan2008}. In this scenario, once off-center ignitions are triggered, deflagration propagates in one direction to break through the stellar surface, then spread rapidly over the stellar surface to collide at the opposite point from the break out point, resulting in a pair of outwardly and inwardly directed jets. Therefore, along the jet axis, the bipolar structure should have the same direction, which is inconsistent with the LOS velocity structure observed in W\,49B. Another possibility is a rapid, differential rotation of a progenitor white dwarf, which is considered to be a possible origin for a superluminous Type Ia SN \citep{Fink2018}. In this scenario, bipolar flows of ejecta developing in opposite directions are expected, and therefore in a fair agreement with our observations.

\subsubsection{Ejecta evolving in bipolar CSM}

Bipolar ejecta flows may alternatively be produced by interaction of initially symmetric ejecta with dense CSM, as was the case for some of the scenarios discussed in \S\ref{sec:equatorial}. To make bipolar flows, the structure of dense CSM needs to be bipolar rather than disk-like or torus-like. 

A simulation geometrically similar to a bipolar CSM case was performed by \cite{Zhou2011}, where a barrel-shaped dense ring around the explosion center was employed. In the simulation, it was shown that ejecta were collimated by the ring and formed an elongated structure along the symmetry axis of the ring. A combined effect of the heating of inner ejecta by the reflected shock from the ring and cooling of outer ejecta by mixing with cooler plasma produced by an evaporated cloud from the ring caused an enhanced emission measure at the center. Moreover, the collimation also forces ejecta flows to be aligned to the ring's symmetry axis, making bipolar kinematics qualitatively similar to those observed with XRISM. The choice of the barrel-shaped ring was motivated by the coaxial rings of CSM observed in the infrared and radio wavelengths, whose physical origin was speculated to be bipolar winds from a massive progenitor star \citep{Keohane2007, Lacey2001}. However, recent abundance measurements rather suggested a Type Ia origin for W\,49B \citep[e.g.,][]{Zhou2018, Sato2025}.

We propose that such a dense, bipolar CSM may be realized by a recently formed planetary nebula \citep{Court2024}. Indeed, such a CSM structure, viewed from the polar direction, was discovered in the SNR N\,103\,B as double rings \citep{Yamaguchi2021}. If a similar CSM structure existed for a progenitor of W\,49B, it may have shaped bipolar flows of hot plasma as the ejecta expanded inside the bipolar cavity, and formed the bar-like X-ray morphology in the edge-on view as simulated by \cite{Zhou2011}. If this is indeed the case, then existence of a binary companion star to the progenitor of W\,49B may be suggested because binary interactions are considered to be a preferred channel to form a bipolar planetary nebula \citep{DeMarco2009, Jones2017}.

A bipolar CSM may alternatively be realized in a CC scenario by a massive progenitor. For instance, a luminous blue variable may form a bipolar CSM, and even if its explosion is spherical, a jet-like, bipolar structure of the ejecta may be developed, as simulated by \citet{Ustamujic2021}. Wolf-Rayet stars may also develop bipolar CSM \citep{Meyer2021}. With a massive progenitor ($\gtrsim 30~M_{\sun}$), a large wind-blown bubble of $\gtrsim 20$~pc is likely formed during its main-sequence phase. The small size of $\approx 5$~pc of the wind-blown bubble suggested in W\,49B was previously argued to be the evidence against such a very massive progenitor \citep{Zhou2018}, while the existence of the wind-blown bubble in this remnant is still debated \citep{Siegel2020}.

The CSM cavity as the primary origin of the bipolar flows is also consistent with the commonality of the LOS velocity structures between elements (\S\ref{sec:elemdep}), although the intrinsic elemental distribution in the ejecta may still cause some differences \citep{Ustamujic2021}. The bipolar CSM may have caused the rarefaction of the plasma by the same mechanism as \citet{Itoh1989}, which may also explain the recombining plasma. Indeed, in the simulation by \cite{Zhou2011}, the inner ejecta that were once hot and dense due to the reflection shock underwent rapid adiabatic expansion in the later stage of evolution, which, in combination with mixing of evaporated cloud materials, made an overionized charge-state distribution. 

\section{Summary} \label{sec:summary}

In this work we have presented the first-ever high-resolution X-ray spectroscopy for the Galactic SNR W\,49B using the Resolve microcalorimeter spectrometer on XRISM. The spectrum is full of emission lines including high-shell transition lines and dielectronic recombination satellite lines, which are the characteristics of a recombining plasma. 

We have investigated the LOS velocity structure of the ejecta using strong lines from various elements. The Fe He$\alpha$ lines are sufficiently bright to perform the pixel-by-pixel velocity measurements. In the pixel map, the overall LOS velocity distribution is characterized by a smooth gradient along the major axis of the ejecta bar, connecting the blueshift in the east and the redshift in the west, with a maximum magnitude of $\sim 300$~km~s$^{-1}$. An asymmetry is found in the gradient between the redshifted and blueshifted parts of the ejecta, indicating a possible influence of the ambient density or an intrinsic inclination difference. The elemental dependence of the LOS velocity structure has been examined with spectra extracted from larger regions dividing the SNR into five parts, finding no significant variations. 

The observed LOS velocities as well as the broadening are inconsistent with an expanding disk model, which is expected for most of the scenarios that assume a spherical explosion. An exceptional case would be one exploding and evolving in a dense CSM with a bipolar structure, which may be expected, for instance, from a Type Ia SN that exploded inside a bipolar planetary nebula. In such a case where the CSM plays a significant role to shape the bipolar ejecta flows, the observed common velocity structures between elements would also be explained easily. A bipolar SN explosion would also explain the kinematics observed with XRISM, while a bipolar CC SN would not be favored in terms of the elemental abundance pattern, as previously pointed out. The XRISM spectrum presented in this paper is also suitable for accurately measuring the elemental abundances, in particular of the Fe-group elements. Hence, these data could potentially resolve this controversy. That work is left to a future publication.

\begin{acknowledgments}

We thank the anonymous reviewer for the careful review and constructive comments. Part of this work was performed under the auspices of the U.S. Department of Energy by Lawrence Livermore National Laboratory under contract DE-AC52-07NA27344. The material is based upon work supported by NASA under award No. 80GSFC21M0002 and by the Strategic Research Center of Saitama University. This work was supported by the JSPS Core-to-Core Program, grant No. JPJSCCA20220002. 

This work was supported by JSPS KAKENHI grant numbers JP23H01211, JP20KK0309, JP21H01136, JP24H00246, JP19K21884, JP20H01947, JP23K20239, JP24K00672, JP24K17093, JP20KK0071, JP22H00158, JP21H01095, JP23K20850, JP21K13963, JP20K14491, JP23H00151, JP21K03615, JP24K00677, JP19K14762, JP23K03459, JP21K13958, JP20H05857, JP23K03454, JP24K17104, JP21H04493, JP24K17105, JP20H01946, JP23K22548, JP20K04009, and JP22H01268, JP23H04899 and by NASA grant numbers 80GSFC21M0002, 80NNSC22K1922, 80NSSC20K0733, 80NSSC24K1148, 80NSSC24K1774, 80NSSC20K0737, 80NSSC24K0678, 80NSSC18K0978, 80NSSC20K0883, 80NSSC25K7064, and 80NSSC18K0988, 80NSSC23K1656.

Lia~Corrales acknowledges support from NSF award 2205918. 
Chris~Done acknowledges support from STFC through grant ST/T000244/1. 
Luigi~Gallo acknowledges financial support from Canadian Space Agency grant 18XARMSTMA. 
Misaki~Mizumoto acknowledges support from Yamada Science Foundation. 
Paul Plucinsky acknowledges support from NASA contract NAS8-0360.
Makoto~Sawada acknowledges the support by the RIKEN Pioneering Project Evolution of Matter in the Universe (r-EMU) and Rikkyo University Special Fund for Research (Rikkyo SFR). 
Atsushi~Tanimoto and the present research are in part supported by the Kagoshima University postdoctoral research program (KU-DREAM). 
Satoshi~Yamada acknowledges support by the RIKEN SPDR Program. 
Irina~Zhuravleva acknowledges partial support from the Alfred P. Sloan Foundation through the Sloan Research Fellowship.

\end{acknowledgments}

\bibliography{ms}{}
\bibliographystyle{aasjournal}



\end{document}